# A way to measure electron spin-flipping at F/N interfaces and application to Co/Cu.

B. Dassonneville, R. Acharyya, H.Y.T. Nguyen, R. Loloee, W.P. Pratt Jr., and J. Bass
Department of Physics and Astronomy, Michigan State University, East Lansing, MI 48823

We describe a technique, using the current-perpendicular-to-plane (CPP) geometry, to measure the parameter $\delta_{F/N}$, characterizing flipping of electron spins at a ferromagnetic/non-magnetic (F/N) metallic interface. The technique involves measuring the CPP magnetoresistance of a sample containing a ferromagnetically coupled $[F/N]_n$ multilayer embedded within the 20 nm thick central Cu layer of a symmetric Py-based, double exchange-biased spin-valve. To focus on $\delta_{F/N}$, the F- and N-layers are made thin compared to their spin-diffusion lengths. We test the technique using F/N = Co/Cu. Analysing with no adjustable parameters, gives inconsistency with $\delta_{Co/Cu} = 0$, but consistency with our prior value of $\delta_{Co/Cu} = 0.25 \pm 0.1$. Taking $\delta_{Co/Cu}$ as adjustable gives $\delta_{Co/Cu} = 0.33^{+0.03}_{-0.08}$.

One of the few remaining fundamental questions about transport of electron spins in ferromagnetic/non-magnetic (F/N) metallic multilayers in the current-perpendicular-to-plane (CPP) geometry is: 'How strong is spin-flipping at F/N interfaces?' In this letter we present and test a general technique for answering this question. To motivate both, we start with some background.

Ref. [1] gave a general technique for measuring the spin-flipping probability at non-magnetic (N1/N2) interfaces, $P_{N1/N2}$, and its related parameter $\delta_{N1/N2}$ given by $P_{N1/N2} = 1 - \exp(-\delta_{N1/N2})$. That technique has been used to obtain values of $\delta_{N1/N2}$ for a number of sputtered N1/N2 pairs [1,2]. It involves inserting an $[N1/N2]_n$ multilayer ($n$ = number of repeats) into the middle of the central Cu layer of a Permalloy (Py = $Ni_{1-x}Fe_x$ with x ~ 0.2) based exchange-biased spin-valve (EBSV). In this EBSV, one Py layer is pinned and the other is free to rotate in small fields to give the parallel (P) and anti-parallel (AP) orientations of Py moments needed for CPP-magnetoresistance (MR) studies. $\delta_{N1/N2}$ is derived by using the theory of Valet and Fert [3] to fit measurements of $A\Delta R = A(R_{AP} - R_P)$ vs $n$, where A is the area through which the CPP current flows and R is the CPP resistance. Unfortunately, simply inserting an F/N multilayer into such a spin valve doesn't work, since it fundamentally alters the CPP-MR in more complex ways. Moreover, $A\Delta R$ for simple $[F/N]_n$ multilayers is relatively insensitive to $\delta_{F/N}$.

Absent a general technique for measuring $\delta_{F/N}$, values for a few F/N pairs have been inferred [4-7] from data not optimized for $\delta_{F/N}$. As explained in [2], we don't trust most of these values. The only one in which we have some trust is our own $\delta_{Co/Cu} = 0.25 \pm 0.1$ [4]. But the validity of even this value was unclear [2].

What was needed is a way to systematically increase the number of F/N interfaces in a sample with an EBSV-like geometry that is sensitive to $\delta_{F/N}$. Our solution is to use a ferromagnetically coupled $[F/N]_n$ multilayer as the central 'moment' of a double EBSV (2EBSV) with pinned, as-identical-as-possible Py layers located symmetrically on both sides. (We note in passing that the technique can also be used with F1/F2 multilayers.) If the F and N layers are each much thinner than their spin-diffusion lengths, and $\delta_{F/N}$ is large enough, then spin-flipping at the F/N interfaces should dominate the variation of $A\Delta R$ with $n$, allowing $\delta_{F/N}$ to be isolated. Using such a symmetric 2EBSV gives good P and AP states, doubles the signal of interest, and simplifies the numerical calculation of $A\Delta R$, which has to be done for only half of the symmetric sample. Applying this technique to F/N = Co/Cu lets us test both the technique and our uncertain value of $\delta_{Co/Cu}$ [4].

Our samples are sputtered in an ultra-high-vacuum-compatible chamber, with 6 sputtering targets, and techniques described in refs. [8,9]. To obtain a uniform CPP-current flow, we use the crossed-superconducting strip geometry, where the multilayer sample of interest is sandwiched between two ~ 1.1 mm wide, 150 nm thick, crossed Nb strips [8,9]. This use limits our measurements to 4.2K. Our samples have the following form (with layer thicknesses in nm):
Nb(150)/Cu(10)/FeMn(8)/Py(6)/Cu(10)/[Co(3)/Cu(1.5)]$_n$/Co(3)/Cu(10)/Py(6)/FeMn(8)/Cu(10)/Nb(150). Here the lower 10 nm of Cu, between the bottom Nb strip and the first FeMn layer, helps the FeMn to grow in the proper structure for pinning the Py, 8 nm of FeMn gives good pinning [10], and the two Cu(10) layers sandwiching the [Co/Cu]$_n$/Co multilayer are thick enough to eliminate exchange coupling between the $[F/N]_n$ multilayer and the Py layers. The [Co(3)/Cu(1.5)]$_n$/Co(3) multilayer gives a sample fully symmetric about its middle. The Py layers are pinned by heating the sample to 453K, applying a magnetic field ~ 200 Oe, and cooling in the field. In magnetization measurements on separately sputtered [Co/Cu]$_n$ multilayers, Cu thicknesses of $t_{Cu}$ = 1.3 and 1.5 nm gave the smallest saturation field, consistent with the ferromagnetic coupling found in [11]. The moment of a [Co/Cu]$_n$/Co multilayer with these thicknesses should then switch as a single entity. Most samples have $t_{Cu}$ = 1.5 nm. But two 'extreme' samples ($n$ = 1 and 8) with $t_{Cu}$ = 1.3 nm give similar results.

We show first that such a 2EBSV gives good P and AP states. Fig. 1 contains an R(H) sweep of magnetic field H from large – H to large + H for a multilayer with $n$ = 1. The moments of the multilayer and the two Py layers are all initially oriented parallel (P) to each other at negative H. At + H ~ 200 Oe the multilayer moment reverses to anti-parallel (AP) to those of the two Py layers. Finally, at H > 1 kOe, the pinning of the two Py layers is overcome, and their moments reverse to parallel (P) to that of the multilayer. Note that the values of R are closely the same for the two different P-states, and that $\Delta$R for $n$ = 1 is already large enough so that we can measure it with good accuracy. We will see below that the scatter in A$\Delta$R for separate 'nominally identical' samples is often larger than the uncertainty implied by Fig. 1. We attribute this scatter to real differences in our complex samples that are not completely under our control. Our specified uncertainties in $\delta_{Co/Cu}$ take account of this larger scatter.

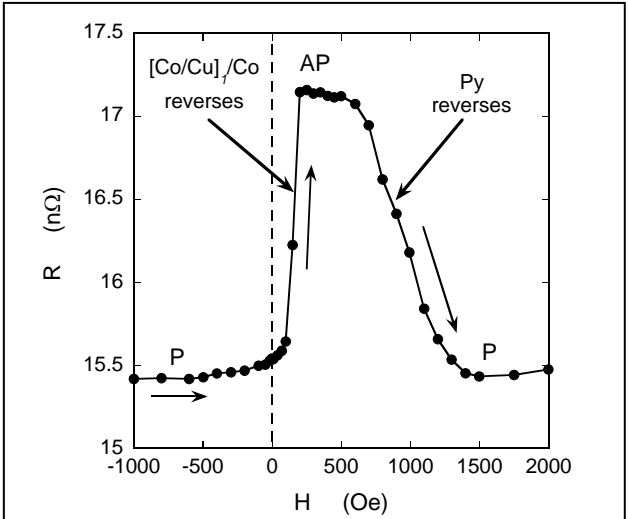

Fig. 1. Sweep of R(H) from − H to + H for a Py-based double exchange-biased spin-valve with a ferromagnetically coupled [Co(3nm)/Cu(1.5nm)]$_{n=1}$/Co(3nm) multilayer in the middle of the central 20 nm thick Cu layer. P and AP indicate states where the moment of the multilayer is parallel (P) or anti-parallel (AP) to those of the two Py layers. Arrows indicate the direction of field sweep.

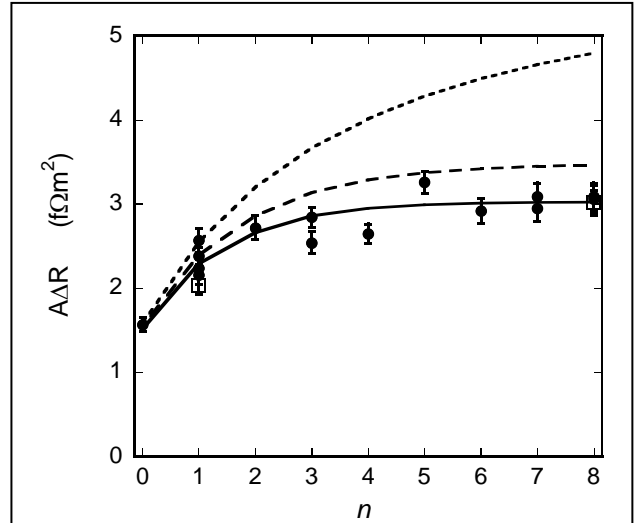

Fig. 2. A$\Delta$R vs $n$ for Py-based 2EBSVs with ferromagnetically coupled [Co(3)/Cu($t_{Cu}$) ]$_n$/Co(3) inserts with $n$ repeats. Filled circles are $t_{Cu}$ = 1.5 nm, open squares are $t_{Cu}$ = 1.3 nm. The short-dashed curve that falls well above the data for $n >1$ is a no-free-parameter calculation assuming $\delta_{Co/Cu}$ = 0. The dashed curve is a similar calculation with the published 'best value' of $\delta_{Co/Cu}$ = 0.25. The solid curve is a fit with $\delta_{Co/Cu}$ adjustable. It gives $\delta_{Co/Cu}$ = 0.33.

To simplify our data analysis, and to ensure that the contributions from $\delta_{Co/Cu}$ dominate, the fixed Co and Cu layer thicknesses in the multilayer were made small compared to their spin-diffusion lengths ($l_{sf}^{Co}$ = 60 ± 20 nm [2,12] and $l_{sf}^{Cu}$ = 500 ± 100 nm [2]). We also fixed all of the parameters in the sample except $\delta_{Co/Cu}$ at values that we previously published. For completeness, we list here these parameters, which are from [9,13], except where explicitly noted: $\beta_{Co}$ = 0.46; $\rho_{Co}^{*}$ = 75 n$\Omega$m; $AR_{Co/Cu}^{*}$ = 0.51 f$\Omega$m$^2$; $\gamma_{Co/Cu}$ = 0.77; $\beta_{Py}$ = 0.76 [14]; $\rho_{Py}$ = 123 n$\Omega$m; $l_{sf}^{Py}$ = 5.5 nm; $AR_{Py/Cu}^{*}$ = 0.50 f$\Omega$m$^2$; $\gamma_{Py/Cu}$ = 0.7; $\rho_{FeMn}$ = 875 n$\Omega$m; $AR_{FeMn/Py}$ = 1.0 f$\Omega$m$^2$; $AR_{Nb/FeMn}$ = 1.0 f$\Omega$m$^2$; $\rho_{Cu}$ = 4.5 n$\Omega$m, and we assumed that spin-flipping is very strong at the FeMn interfaces with Py [1].

Turning now to our data, Fig. 2 shows a plot of A$\Delta$R vs $n$, which we analyse numerically using the theory of Valet and Fert [3]. The dotted curve is a no-free-parameters calculation of A$\Delta$R assuming $\delta_{Co/Cu}$ = 0. This curve fits the data for $n$ = 0, confirming that our other parameters are okay, but lies increasingly far above the data as $n$ grows. Clearly, $\delta_{Co/Cu}$ = 0 won't do. The dashed curve is also a no-free-parameters calculation, but now including our previously published 'best value' of $\delta_{Co/Cu}$ = 0.25. This curve falls close to our new data, but slightly above it. The new data of Fig. 2 require a slightly larger $\delta_{Co/Cu}$. To determine a 'best value' of $\delta_{Co/Cu}$ from these new data, we took the average of A$\Delta$R from $n$ = 6-8 and required the calculated value of A$\Delta$R at $n$ = 7 to agree with this value, adjusting only $\delta_{Co/Cu}$. The result is the solid curve, with $\delta_{Co/Cu}$ = 0.33. Note that the curve fits the rest of the data rather well. The value 0.33 lies within the uncertainty of our previous estimate: $\delta_{Co/Cu}$ = 0.25 ± 0.1 [4], supporting both that value and the new technique. To determine a final 'best estimate' of $\delta_{Co/Cu}$ and its uncertainties, we consider both effects of slightly varying other parameters within their uncertainties, and what we can learn from earlier data sets collected together in Figs. 13 and 14 of ref. [4].

Letting $l_{sf}^{Co}$ vary by ± 20 nm, or $l_{sf}^{Cu}$ by ± 100 nm, changes $\delta_{Co/Cu}$ by only ~ 0.01. Taking $l_{sf}^{Co}$ down to 9 nm can fit the data of Fig. 2 with $\delta_{Co/Cu}$ = 0. But such a short $l_{sf}^{Co}$ is incompatible with data in refs. [12] and [15]. We can slightly improve the fit to the average value of the $n$ = 1 data in Fig. 2 by increasing $\gamma_{Co/Cu}$ from 0.77 to 0.78, within its uncertainty of ± 0.04 [9]. Such an increase gives a best fit of $\delta_{Co/Cu} = 0.34$. From Fig. 13 in ref. [4], the 'best fit' to the ratio $(A\Delta R)_{sep}/(A\Delta R)_{int}$ for separated vs interleaved Co/Cu multilayers was $\delta_{Co/Cu} \sim 0.35$, near our present 'best value'. $\delta_{Co/Cu} \sim 0.35$ also gives a better fit than 0.25 to the asymmetric Co/Cu EBSV data (open symbols) in Fig. 14 of ref. [4] [Note: the error bar for the asymmetric curve there is incorrect—it should be about three times longer]. However, $\delta_{Co/Cu} \sim 0.35$ gives a worse fit to the symmetric spin-valve data in Fig. 14 of ref. [4].

Taking into account all of the data in both the present paper and ref. [4], we arrive at a best estimate of $\delta_{Co/Cu} = 0.33^{+0.03}_{-0.08}$.

To summarize, we have presented a technique to determine the interfacial spin-flipping parameters $\delta_{F/N}$ for F/N pairs (or $\delta_{F1/F2}$ for F1/F2 pairs), when the F and N (or F1 and F2) spin-diffusion lengths are not too short, and where multilayers can be made to have ferromagnetic coupling. The technique involves embedding a ferromagnetically coupled $[F/N]_n$ (or $[F1/F2]_n$) multilayer in the middle Cu layer of a symmetric Py-based double exchange-biased spin-valve, and measuring the change in specific resistance, $A\Delta R = A(R_{AP} - R_P)$ between the parallel (P) and anti-parallel (AP) alignments of the multilayer and the Py moments as a function of bilayer number $n$. In the present paper, we applied this technique to $[Co/Cu]_n$ multilayers, and found the following coupled results. First, the data are inconsistent with $\delta_{Co/Cu} = 0$. Second, a no-free-parameters calculation using a previously published 'best estimate' of $\delta_{Co/Cu} = 0.25$ comes close to the data. Third, a 'best fit' of $\delta_{Co/Cu} = 0.33^{+0.03}_{-0.08}$ falls within the uncertainty range of the previously published value. We conclude that our results strongly suggest that our technique is reliable, and confirm a non-zero value for $\delta_{Co/Cu} \approx 0.3$.

We hope that development of this technique, and solidification of the results for Co/Cu, will stimulate more theory to establish the source(s) of spin-flipping at F/N interfaces. Likely contributions include spin-orbit scattering and interfacial spin-disorder [16,17]. A fundamental issue not yet resolved is how sensitive $\delta_{F/N}$ is to changes in interfacial structure. We can say the following about this issue. (1) Studies by resistance, x-rays, and TEM [18,19] show that our sputtered interfaces are intermixed over a few monolayers. Since our CPP-MR data are reproducible under random variations in interfacial structure from sample to sample, $\delta_{Co/Cu}$ does not seem to be sensitive to precise details of such variations. (2) From a combination of experiment and theory, Ref. [20] shows that the other basic interface parameter, the interface specific resistance, is often only weakly sensitive to intermixing when the two metals are close to lattice matched, as are Co and Cu. From these results, we conclude that the sensitivity of $\delta_{F/N}$ to interfacial intermixing and roughness is not yet known. Since it is difficult to both controllably vary and characterize interfacial structure experimentally, theoretical analysis of effects of different interfacial structures on $\delta_{F/N}$ would be helpful.

Acknowledgements:. This research was supported in part by NSF grant DMR-08-04126.